# Email as Spectroscopy: Automated Discovery of Community Structure within Organizations

Joshua R. Tyler, Dennis M. Wilkinson, Bernardo A. Huberman
HP Labs, 1501 Page Mill Road, Palo Alto CA, 94304
*{jtyler; dennisw; huberman}@hpl.hp.com*

**Abstract.** We describe a methodology for the automatic identification of communities of practice from email logs within an organization. We use a betweenness centrality algorithm that can rapidly find communities within a graph representing information flows. We apply this algorithm to an email corpus of nearly one million messages collected over a two-month span, and show that the method is effective at identifying true communities, both formal and informal, within these scale-free graphs. This approach also enables the identification of leadership roles within the communities. These studies are complemented by a qualitative evaluation of the results in the field.

# Introduction

Email has become the predominant means of communication in the information society. It pervades business, social and technical exchanges and as such it is a highly relevant area for research on communities and social networks. Not surprisingly, email has been established as an indicator of collaboration and knowledge exchange [Wellman, 2002, Whittaker & Sidner, 1996]. Email is also a tantalizing medium for research because it provides plentiful data on personal communication in an electronic form. This volume of data enables the discovery of shared interests and relationships where none were previously known [Schwartz & Wood, 1992]. Given its ubiquity, it is a promising resource for tapping into the dynamics of information within organizations, and for extracting

the hidden patterns of collaboration and leadership that are at the heart of communities of practice.

Communities of practice are the informal networks of collaboration that naturally grow and coalesce within organizations. Any institution that provides opportunities for communication among its members is eventually threaded by communities of people who have similar goals and a shared understanding of their activities [Ouchi, 1980]. These communities have been the subject of much research as a way to uncover the structure and communication patterns within an organization—the reality of how people find information and execute their tasks. (for example, see [Blau and Scott, 1962], [Burt, 1980], or for a survey see [Scott 1992]).

These informal networks coexist with the formal structure of the organization and serve many purposes, such as resolving the conflicting goals of the institution to which they belong, solving problems in more efficient ways [Huberman & Hogg, 1995], and furthering the interests of their members. Despite their lack of official recognition, informal networks can provide effective ways of learning, and with the proper incentives actually enhance the productivity of the formal organization [Crozier, 1964, Crane, 1972, Lave & Wenger, 1991].

Recently, there has been an increased amount of work on identifying online communities (a brief overview of this work can be found in [Wellman, 2001]). Wellman finds that online relationships do indeed reflect actual social relationships thus adding effectively to the social capital [Wellman, 2002]. Mailing lists and personal web pages also serve as proxies for social relationships [Adamic & Adar, 2002], and the communities identified from these online proxies resemble the actual social communities of the represented individuals. Ducheneaut and Bellotti conducted in-depth field studies of email behavior, and found that membership in email communities is quite fluid and depends on organizational context [Ducheneaut & Bellotti, 2002].

Because of the demonstrated value of communities of practice, a fast, accurate method of identifying them is desirable. While the literature pertaining to the theory of communities of practice is extensive, there is less work devoted to the task of identification. Classical practice is to gather data from interviews, surveys, or other fieldwork and to construct links and communities by manual inspection (see [Allen, 1984], [Hinds & Kiesler, 1995] or an Internet-centric approach in [Garton, Haythornwaite & Wellman, 1999]). These methods are accurate but time-consuming and labor-intensive, prohibitively so in the context of a very large organization. Alani et al. [Alani et al., 2002] recently introduced a semi-automated utility that uses a simple algorithm to identify nearest neighbors to one individual within a university department. However, this program relies on previously collected relational data that may be difficult to obtain for a given organization.

This paper presents a fully automated method for identifying communities of practice within an organization. The method uses email data to construct a network of correspondences, and then discovers the communities by partitioning this network in a particular way, which we describe below. The only pieces of information used from each email are the names of the sender and receiver (i.e., the "to:" and "from:" fields), enabling the processing of a large number of emails while minimizing privacy concerns.

We describe an experiment performed within our own organization, HP Labs, using nearly one million email messages collected over a period of roughly two months. The method was able to identify small communities within this 400-person organization in a matter of hours, running on a standard Linux desktop PC. In addition, we utilized the network of correspondence to identify leadership within these communities. This experiment was followed by a qualitative evaluation of the experimental results in the "field", which consisted of sixteen face-to-face interviews with individuals in HP Labs. These interviews validate the results obtained by our automated process, and provide interesting perspectives on the communities identified.

## Identifying Communities

The method we used to automatically identify communities within an organization consists of two basic steps. The first one uses the headers of email logs to construct a graph where the vertices are senders or recipients of email messages and the links denote a direct email between the nodes they connect. The second step uses the algorithm that we describe below to find the communities embedded in the graph.

It is straightforward to construct a graph based on email data, in which vertices represent people and edges are added between people who corresponded through email. The minimum number of messages passed between any two vertices defines the threshold that one can vary to construct the graph. We find that graphs created in this way are power-law for high threshold values, in the sense that there are a few vertices with a high number of links and many with few links. Given that our measured exponent is less than 3.5, we expect the graphs to consist of one giant connected component and many smaller isolated components of O(1) vertices [Aiello et al., 2000]. Since the smaller components can clearly be identified as communities; the task remains to identify communities within the giant component.

A graph can be said to have community structure if it consists of subsets of vertices, with many edges connecting vertices of the same subset, but few edges lying between subsets [Girvan & Newman, 2002]. Finding communities within a graph is an efficient way to identify groups of related vertices. We applied the non-local process of Wilkinson and Huberman [Wilkinson & Huberman, 2002] which partitions a graph into discrete communities of nodes and is based on the idea of betweenness centrality, first proposed by Freeman [Freeman, 1979]. A key feature of this process is that, based on the structure of the graph, it is able to distinguish and suppress isolated inter-community correspondences, so that the correspondents involved are placed in different communities.

In order to explain the community discovery process we consider as a first example the small graph shown in Figure 1. This graph consists of two well-defined communities: the four vertices denoted by squares, including vertex A, and the nine denoted by circles, including vertex B.

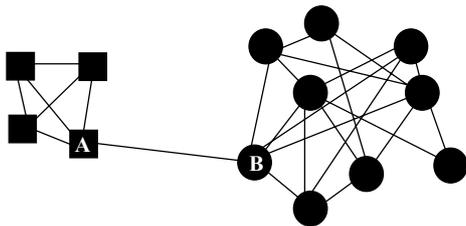

**Figure 1. An example graph illustrating betweenness.**

In the context of Figure 1, edge AB has the highest betweenness. If we were to remove it, the graph would split into two connected components, the square and circle communities. This illustrates the idea behind our method of imposing community structure on a graph: we repeatedly identify inter-community edges of large betweenness and remove them, until the giant component is resolved into many separate communities.

To find inter-community edges, we exploit Freeman's [Freeman, 1979] notion of betweenness centrality, or betweenness, applied to edges. The betweenness of an edge is defined as the number of shortest paths that traverse it. This property distinguishes inter-community edges, which link many vertices in different communities and have high betweenness, from intra-community edges, whose betweenness is low.

The removal of an edge strongly affects the betweenness of many others, and so we must repeatedly recalculate the betweenness of all edges. To do this quickly, we used the fast algorithm of Brandes [Brandes, 2001, Newman, 2001, Girvan and Newman, 2002], whose basic strategy is the following: Consider the shortest paths between a single vertex, the "center", and all other vertices. Calculate the betweenness of each edge due only to these shortest paths, and add them to a running total. Then change centers and repeat until every vertex has been the center once. The running total for each edge is then equal to exactly twice the exact betweenness of that edge, because we have considered all the pairs of endpoints of paths twice.

Our procedure stops removing edges when we cannot further meaningfully subdivide our communities; for example, as in Figure 1, after removing edge AB. What criterion tells us when to stop? As we remove edges, we divide the graph into many unconnected components. Structurally, a component of 5 or fewer vertices cannot consist of two viable communities. The smallest possible such component is size 6, consisting of two triangles linked by one edge (Figure 2). If at any time we remove an edge from our graph and separate a component of size < 6, we can identify it as a community.

Components of size ≥ 6 can also be individual communities, like the group of 9 in Figure 1. To identify this type of component as a community, we use an intuitive threshold based on the betweenness of an edge connecting a <u>leaf</u> vertex, or vertex of degree one, to the rest of the graph. Consider the graph of Figure 3 below. It is clear that it consists of just one community. Applying the Brandes algorithm, we find that edge XY has the highest betweenness, indicating that the size of the largest distinct community within the graph has size 1. That is, there are no distinct communities within the graph.

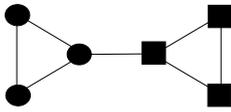

**Figure 2. The smallest possible graph of two viable communities.**

In general, the single edge connecting a leaf vertex (such as X in Figure 3) to the rest of a graph of $N$ vertices has a betweenness of $N-1$, because it contains the shortest path from X to all $N-1$ other vertices. The stopping criterion for components of size ≥ 6 is therefore that the highest betweenness of any edge in the component be equal to or less than $N-1$.[1]

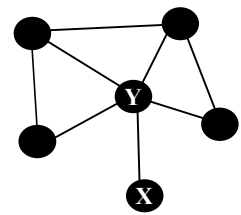

**Figure 3. An example graph of one community that does not contain distinct sub-communities.**

## Multiple Community Structures

As mentioned above, the removal of any one edge affects the betweenness of all the other edges. Therefore, the order of removal of edges affects which edges are removed, particularly in large, real-world graphs such as the email graph. Early in the process, there are many inter-community edges which have high betweenness and the choice of which to remove, while arbitrary, dictates which edges will be removed later. We take advantage of this arbitrariness to repeatedly partition the graph into many different sets of communities. We then compare the different sets and aggregate the result into a final list of communities.

Identifying many community structures, in the form of different sets of communities, and aggregating them into a final list provides a much richer result than simply partitioning the graph once. Consider the placement of John and Sarah in communities. If John appears within the same community in all 50 sets, it is clear that John definitely belongs to that community. The order of edge removal had no effect on him. However, if Sarah appears in one community in some sets in another (or even several others) in other sets, the order of edge removal did affect her and we should consider that she has some affiliation with those two (or more) communities. Had we only considered one community structure, Sarah would have been placed in one community, rather arbitrarily, and we would have lost information about her role in the other community (communities).

---

[1] It is not in general true that an inter-community edge must have betweenness greater than N-1. For a community of size m within a graph of size N, there is a total betweenness of $m(N-m)$ divided among the edges connecting the community to the graph. So, if there are more than m such edges, it is possible that none of them will have betweenness greater than N. However, remember that none of these edges, or the extra-community vertices they connect, should be adjacent, because then m would not be a community. This type of situation is extremely unlikely in a power law graph of degree ~ 2, such as the email graph.

The small graph of Figure 4 illustrates the mechanics of the placement of a vertex such as Sarah (vertex B in the graph) into a community. The graph consists of two communities, one on the left including vertex A, and another on the right including C. Among its edges, BC initially has the highest betweenness, and AB's betweenness is also high. If we choose to remove BC first, AB becomes an intra-community edge with low betweenness which will never be removed, and vertex B will eventually be placed in a community with vertex A. Had we removed AB first, BC would have been rendered intra-community, and vertex B would end up in the community with C.

Moreover, in considering Figure 4, it is not clear where B should end up. In fact, from the graph alone, B could rightfully be considered to be a part of both communities. The choice of one or the other is made early in the process, arbitrarily, when either edge BC or edge AB is removed.

However, the arbitrary nature of this choice is a help rather than a hindrance if we use it to partition the graph repeatedly into different, plausible structures. We introduce the randomness into our algorithm in the following way. In partitioning a large connected component of vertices, instead of using every vertex as the "center" once, we cycle randomly through at least m centers (where m is some cutoff) until the betweenness of at least one edge exceeds a threshold, again based on the betweenness of a "leaf" vertex. We then remove the edge whose betweenness is highest at that point, and repeat until we have broken the graph into communities, identified by the criterion described above. We only do this for large components, using the full Girvan-Newman algorithm for small ones. The modified algorithm may occasionally remove an intra-community edge, but such errors are unimportant when one compares a large number of community structures.

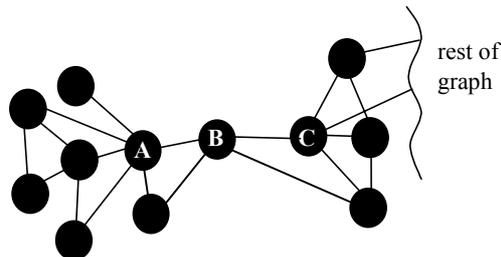

**Figure 4. An example of a denser community subgraph. Ambiguity is introduced by the sequence in which edges are removed.**

Applying this modified process n times, we obtain n community structures imposed on the graph. We can then compare the different structures and identify communities. For example, after imposing 50 structures on our graph, we might find: a community of people A, B, C, and D in 25 of the 50 structures; a community of people A, B, C, D, and E in another 20; and one of people A, B, C, D, E and F in the remaining 5. We report this result in the following way:

A(50) B(50) C(50) D(50) E(25) F(5)

which signifies that A, B, C, and D form a well-defined community, E is related to this community, but also to some other(s), and F is only slightly, possibly erroneously, related to it. For details of the aggregation procedure, please see [Wilkinson & Huberman, 2002].

The entire process of determining community structure within the graph is displayed in Table I.

A. For i iterations, repeat {
    1. Break the graph into connected components.
    2. For each component, check to see if component is a community.
        a. If so, remove it from the graph and output it.
        b. If not, remove edges of highest betweenness, using the modified Brandes algorithm for large components, and the normal algorithm for small ones. Continue removing edges until the community splits in two.
    3. Repeat step 2 until all vertices have been removed from the graph in communities.
}
B. Aggregate the i structures into a final list of communities.

**Table I. Algorithm for determining community structure**

# Results

We performed an experiment of our algorithm using email data from the HP Labs mail server[2]. Starting from an original set of 878,765 logged emails over the period 25 November 2002 to 18 February 2003, we constructed a "clean" subset of 185,773 emails between any two of the 485 current HP Labs employees. For reasons of privacy and complexity, we neglected emails that had an external origin or destination. We also excluded messages sent to a list of more than 10 recipients, as these emails were often lab-wide announcements (rather than personal communication), which were not useful in identifying communities of practice.

From this data, we created a graph as described in the previous section. We used a threshold of 30 messages - that is, a pair was linked by an edge in the graph if and only if the two individuals had exchanged at least 30 total emails in our dataset, and each had sent at least 5 emails to the other (to reduce the number of one-way relationships). This threshold excluded some individuals who either sent very few emails or used other email systems. The graph thus created consisted of 367 nodes, connected by 1110 edges.

---

[2] This experiment was performed using a 900 MHz PC (1 GB of RAM) running the Linux operating system, and ran for approximately two hours.

This graph was power law in degree, as mentioned above, with exponent 3.15 (Figure 5). It consisted of one giant connected component of 343 nodes and many smaller components ranging in size from 2 to 8 (Table II). We applied our algorithm as described above to identify the communities within the graph.

| size of component | number of such components |
|---|---|
| 343 | 1 |
| 8 | 1 |
| 4 | 2 |
| 3 | 2 |
| 2 | 1 |

Table II. Connected component sizes in email graph, threshold 30

66 distinct communities were detected, including the small components. The largest community consisted of 57 individuals, and there were several communities of size 2. The mean community size was 8.4, with standard deviation 5.3. Comparing these communities with information from the HP corporate directory, we found that 49 of the 66 communities consisted of individuals entirely within one lab or organizational unit. The remaining 17 contained individuals from two or more organizations within the company.

We demonstrate the form of our results in Table III, which shows a sample community from the list produced by applying our method to the HP Labs email graph. The login names have been disguised for privacy; individuals have been randomly assigned an identification number. We performed 50 iterations of the modified Brandes algorithm described above to partition the graph into 50 different structures, and aggregated these structures into a list of communities. "Strength in community" is a count of how many times an individual belonged to a particular community after the graph was partitioned. In this particular case, five people were placed in the community in every iteration, while three others were only sometimes grouped into the community.

| Example community | |
|---|---|
| Individual | Strength in community (max 50) |
| Person 34 | 50 |
| Person 267 | 50 |
| Person 56 | 50 |
| Person 406 | 50 |
| Person 212 | 50 |
| Person 246 | 29 |
| Person 331 | 15 |
| Person 87 | 7 |

Table III. An example community from our results

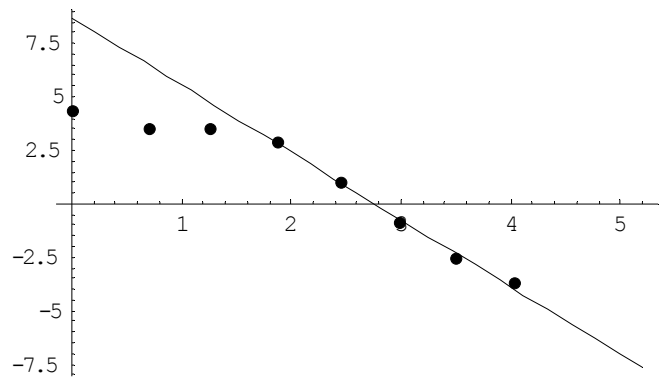

**Figure 5.** The number of vertices (y-axis) is plotted against the degree of the vertex (x-axis) for the email graph, threshold 30, on a log-log scale. We followed a typical exponential binning procedure in plotting the data [Newman, 2003]. The deviation from the power law for low vertex degree is to be expected in a small sample, and has little effect on the topological properties of the graph.

## Identifying Leadership Roles

In addition to formal and informal work communities, it is also possible to draw inferences about the leadership of an organization from its communication data. With the ubiquity of electronic communication, one might expect that it would flatten out the corporate hierarchy—that the structure of an organization would not be visible in, for example, email log data. Sproull and Kiesler, in a relatively early study of email usage, found that email reduces social context cues, such as the relationship between managers and subordinates, and promotes "status equalization" within the medium [Sproull & Kiesler, 1986].

We have found that the structure of an email network bears some similarities to the structure of the organization. We constructed a graph of the email network with a standard force-directed spring algorithm [Fruchterman & Rheingold, 1991], shown in Figure 6[3]. The vertices have repulsive forces between each other, and the edges are springs pulling together the vertices at either end. This spring layout of the email network does not use any information about the actual organization structure. Each vertex represents an individual, and each edge signifies that the total email sent between the two connected vertices exceeds a given threshold (30 messages total in the 2½ months of data, with a minimum of 5 from each).

The graph has also been color-coded—the level of redness corresponds to closeness to the top of the organization hierarchy. There is one totally red vertex (RGB value 255,0,0) representing the lab director. The bluer a vertex is, the more hierarchy levels separate the represented individual from the lab director. The maximum depth in this network is six levels from the lab director.

---

[3] The layout was done with Zoomgraph, a zoomable graph layout tool, available at http://www.hpl.hp.com/shl/projects/graphs/

Visual inspection of the graph reveals the organization leadership tends to end up in the center—the reddest vertices are in the center of the graph. Measuring distance from the center of the graph provides evidence of this trend. Table IV shows the average hierarchy depth (levels from the lab director) for groups at increasing radii from the center. The first set of vertices is all those lying within a circle of radius 0.1 of the center, the second set is those of radius 0.1 to 0.2, and so on, where the height and width of the graph are 1.

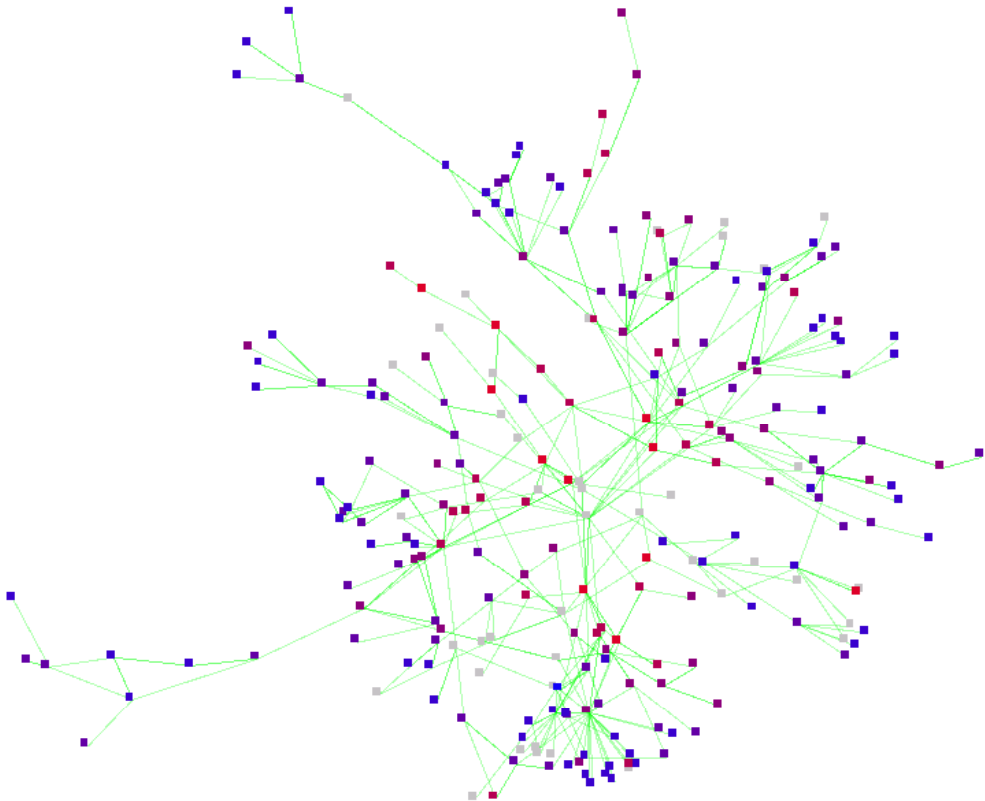

**Figure 6. The giant connected component of the HP Labs email network. The redness of a vertex indicates an individual's closeness to the top of the lab hierarchy (red=close to top, blue=far from top, gray=no data available).**

## Applications

Evaluating communication networks with this technique could provide information about leadership in communities about which little is known. Sparrow proposed this approach for analyzing criminal networks [Sparrow, 1991], noting that "Euclidean Centrality is probably the closest to the reality" of the current criminal network analysis techniques. More recently, Krebs applied centrality measures and graphing techniques [Krebs, 2002] to the terrorist networks uncovered in the 9/11 aftermath. In both of these cases, however, the links between entities were added manually. Our results suggest that meaningful insights about organization leadership can be drawn automatically from a simple record of communication interactions.

## Field Evaluation

In order to validate the results of the communities identified by our algorithm, we conducted a brief, informal field study, consisting of personal interviews with sixteen individuals within our research lab (our experiment's population). In these interviews, we presented the subjects with the community or communities in which they were placed by the algorithm, and then invited them to comment on these results.

| Distance from center | # of vertices | Average hierarchy depth |
| --- | --- | --- |
| < 0.1 | 23 | 2.05 |
| 0.1 to 0.2 | 71 | 3.43 |
| 0.2 to 0.3 | 91 | 3.88 |
| 0.3 to 0.4 | 83 | 4.01 |
| 0.4+ | 51 | 3.66 |

Table IV. Average hierarchy depth by radius from graph center.

### Interview Technique

Interviews were conducted in the subject's office, and lasted from 5-30 minutes, with an average of around 15 minutes. The interview was structured by a few open-ended questions, intended to provoke discussion. First, the subject was asked to characterize the community in which he or she appeared—did it "make sense," and if so, how it could be described, in the subject's own words. Then, once the subject had defined the nature of the community, we asked if there were any people missing from the community, or if there were people in the community who should not have been. Finally, we asked for more general comments on the plausibility of this community.

The interviews were recorded with handwritten notes from two joint interviewers. Following the interviews, these notes were compared and analyzed to identify consistent issues. In this paper, we present some representative quotes from our interviews as a way of illustrating some of our findings. Where there was diversity in the responses, we offer multiple examples representing the various perspectives articulated.

# Results

We interviewed sixteen individuals in seven different communities. The communities were chosen arbitrarily to give us a representation of various community sizes and levels of departmental homogeneity. They ranged in size from four to twelve people, and three out of the seven were heterogeneous (included members of at least two different departmental units within the company).

All sixteen subjects gave positive affirmation that the community reflected reality. More specifically, eleven described the group as reflecting their department, four described it as a specific project group, and one said it was a discussion group on a particular topic.

Nine of the sixteen (56.25%) said nobody was missing from the group, six people (37.5%) said one person was missing, and one person (6.25%) said two people were missing. Conversely, ten of the sixteen (62.5%) said that everybody in the group deserved to be there, whereas the remaining six (37.5%) said that one person in the group was misclassified.

## Departmental Communities

As expected, our interviews confirmed that many, and perhaps most, of the communities identified are based on organization structure. Some responses to seeing the generated communities were "yes, that's my department," "this is a group that reports to me," and "that's pretty much our project team."

However, the communities also tended to include people who were de facto department members, but who did not technically appear in the department's organization chart, such as interns or people whose information was out of date. For example, one person said of his identified community, "One of these was an intern, and the rest are in the same group."

Finally, the algorithm seemed to succeed in dividing departmental groups whose work is distinct, but lumped together groups whose projects overlap. For example, one manager said, "This is one category of my direct reports… I have two groups of reports… The two groups are separate. They don't meet together and they don't work together." By contrast, in our largest group of twelve, the three people we interviewed all informed us that the group comprised their entire department, and that they don't work with all eleven other people. However, one also said, "there tends to be a lot of overlap in the projects in our department."

## Informal Project or Discussion Communities

We were particularly interested in exploring the heterogeneous, cross-department communities that were identified. We found that most of them represented groups formed around specific projects, and in one case, a discussion forum.

In one case, we found a community containing three people from different labs coordinating on one project. This group included a technology transfer project manager, a researcher who was the original designer of a piece of PC hardware, and an engineer redesigning the hardware for a specific printer. As the engineer described it, "This is a printer board project… I'm completely out of this group, organization-wise. They don't have engineers that will build this kind of stuff [he holds up a printed circuit board]."

Another interesting example was that of the discussion-based community. One subject in the group said, "It's interesting that this is not an administrative group. I've been part of a discussion forum on multimedia networking… Your algorithm is good at identifying the people I do my technical work with." She continued to explain that she thought the eight-person community was accurate for her discussion group, even though six of the eight were on a shared email list that the other two were not.

Communities of Practice Have Leaders

In our interviews, we discovered one similarity in all seven communities: The presence of a leader or manager of the group. In the departmental communities, this was the department manager. From the smallest department we identified (four people) to the largest (twelve), the manager was included in the community. In the more informal communities, the subjects identified a person who was the manager or coordinator of the group. In the printer hardware project example described above, this was the technology transfer project manager (which was confirmed by all three interviews in this group).

Another community was formed around work related to a recent merger completed by the company. In this community, one person from the lab director's office joined with five people in the lab finance department, and became the leader of this project.

# Discussion

The power of this method for identifying communities and leadership is in its automation. We have found that it does an effective job of uncovering communities of practice with nothing more than email log ("to:" and "from:") data. Its simplicity and suitability for scale-free networks means that it can be applied to organizations of thousands and produce results efficiently. Furthermore, it also applies to other forms of communication, such as instant messaging, telephony, SMS-style text messaging, and mobile device usage.

This technique is a useful complement to existing strategies for identifying communities of practice. In cases where an organization is very large, widely dispersed, or incompletely defined (informal), this method provides an alternative to the more traditional, labor-intensive approaches. Our empirical test with nearly one million messages from an organization of 400, using only a desktop PC, suggests that it would be simple to scale this analysis up to organizations of many thousands.

As discussed previously, an intriguing application of this technique is in the area of intelligence and covert networks. For these "organizations", which have no official documented structure, the identification of communities and potential leaders in this way, based on communication records, could prove useful. Sparrow [Sparrow, 1991] and Krebs [Krebs, 2002] both explore this concept, but to date there has not been (at least in the public domain) a treatment of a very large informal organization.

## Limitations

As with any evaluation of centrality measures, defining the organization population is important. In a setting like a corporate lab, this is easy—the membership is clearly defined. In an informal network, however, this is much more difficult. In addition, disambiguating redundant names or IDs is a tricky task. Within HP Labs, we were able to use the corporate directory to accomplish this in creating our "clean" dataset, but in an informal network, without such a reference, it would be much harder.

Since a computer generates them, the communities we identify lack the richness in contextual description provided by ethnomethodological approaches. We do not know the nature or character of the identified communities, the relative importance of one community to another, or the subtle inter-personal dynamics within the communities. These kinds of details can only be uncovered with much more data- or labor-intensive techniques.

Finally, we found that in some cases this algorithm can make mistakes. We found that a few of the communities in our experiment contained misclassified individuals, and some were missing individuals who should have been included.

## Future Work

We are particularly interested in extending our analysis of communities to include a temporal dimension. Our email data also includes a timestamp of when email was sent, enabling us to build communities based only on email from one particular week or one specific time of day (e.g. the "Monday morning" communities). We plan to investigate how community membership changes over time, such as from week to week or from day to day within the week.

We are also improving our understanding of how to set the threshold on which data to use. Currently, we have a finite threshold of 30 messages between a pair of individuals, which neglects the fact that some people send more email overall than others. We are exploring thresholds based on percentages and variances of

email traffic, as well as comparing the communities generated by using different thresholds.

Finally, we are attempting to expand our dataset to include more HP employees outside just HP Labs. Testing our technique on an organization of 150,000 people, or at least a significant portion thereof, will demonstrate whether it scales effectively to very large organizations.

# Acknowledgements

The authors would like to acknowledge the efforts of Andrea Chenu, Charlie Schramm, and Herb Strandberg in helping us obtain the email data, as well as all of our interview subjects.

# References

Adamic, L. A., and Adar, E. (2002). "Friends and Neighbors on the Web," to appear in *Social Networks*.

Aiello, W., Chung, F., & Lu, L. (2000). "A Random Graph Model for Massive Graphs", in *Proceedings of the 32nd Annual ACM Symposium on Theory of Computing*, 171-180.

Alani, H., O'Hara, K., and Shadbolt, N. "ONTOCOPI: Methods and Tools for Identifying Communities of Practice, Intelligent Information Processing Conference", *IFIP World Computer Congress (WCC)*, Montreal, Canada, 2002.

Allen, T. (1984). *Managing the Flow of Technology*. MIT Press.

Brandes, U. (2001) "A Faster Algorithm for Betweenness Centrality", *Journal of Mathematical Sociology* 25(2):163-177.

Burt, R. S. (1980). "Models of Network Structure", *Annual Review of Sociology*, Vol. 6, pp. 79-141.

Crane, D. (1972). *Invisible Colleges: Diffusion of Knowledge in Scientific Communities*. University of Chicago Press, Chicago.

Crozier, M. (1964). *The Phenomenon of Bureaucracy*. University of Chicago Press, Chicago.

Ducheneaut, N. and Bellotti, V. (2002). "A Study of Email Work Processes in Three Organizations," submitted to *The Journal of Computer-Supported Cooperative Work*. Draft furnished by the authors.

Freeman, L. (1977) "A Set of Measures of Centrality Based on Betweenness", *Sociometry* 40, 35-41.

Fruchterman, T. M., and Rheingold, E. M. (1991). "Force-Directed Placement", *Software Experience and Practice*, 21(11).

Girvan, M., & Newman, M. (2002) "Community structure in social and biological networks", *Proc. Natl. Acad. Sci.* USA 99, 8271-8276.

Huberman, B. A. and Hogg, T. (1995). "Communities of Practice: Performance and Evolution". Computational and Mathematical Organization Theory, Vol. 1, pp. 73-92.

Krebs, V. E., (2002). "Uncloaking Terrorist Networks," *First Monday*, volume 7, number 4 (April 2002).

Lave, J. and Wenger, E. (1991). *Situated Learning: Legitimate Peripheral Participation*. Cambridge University Press.

Newman, M., (2001). "Who is the Best Connected Scientist? A study of scientific coauthorship networks", *Phys Rev*. E 64.

Newman, M.E.J. (2003) "The structure and function of complex networks", to appear in *SIAM Review*, June 2003

Ouchi, W. G. (1980). "Markets, Bureaucracies, and Clans." *Administrative Science Quarterly*, Vol. 25, pp. 129-141.

Schwartz, M. F., and Wood, D. C. M. (1993). "Discovering Shared Interests Among People Using Graph Analysis", Communications of the ACM, volume 36, issue 8, pp. 78-89.

Scott, W. R. (1992). *Organizations: Rational, Natural, and Open Systems*. Prentice-Hall, Inc., Englewood Cliffs, New Jersey.

Sparrow, M. K. (1991). "The Application of Network Analysis to Criminal Intelligence: An Assessment of the Prospects," *Social Networks,* volume 13, pp. 251-274.

Sproull, L., and Kiesler, S. (1986). "Reducing social context cues: Electronic mail in organizational communication", *Management Science*, 32, 1492-1512.

Wellman, B. "Computer Networks As Social Networks", *Science* 293, 14, Sept 2001: 2031-34.

Whittaker, S., and Sidner, C. (1996). "Email Overload: Exploring Personal Information Management of Email", in *Proceedings of CHI '96*, ACM Press, 276-283.

Wilkinson, D. and Huberman, H (2002). "A Method for Finding Communities of Related Genes", submitted for publication, http://www.hpl.hp.com/shl/papers/communities/index.html.